\documentclass[pdflatex,sn-mathphys-num]{sn-jnl}% Math and Physical Sciences Numbered Reference Style
\usepackage{graphicx}%
\usepackage{multirow}%
\usepackage{amsmath,amssymb,amsfonts}%
\usepackage{amsthm}%
\usepackage{mathrsfs}%
\usepackage[title]{appendix}%
\usepackage{xcolor}%
\usepackage{textcomp}%
\usepackage{manyfoot}%
\usepackage{booktabs}%
\usepackage{algorithm}%
\usepackage{algorithmicx}%
\usepackage{algpseudocode}%
\usepackage{listings}%
\usepackage{url}
\usepackage{hyperref}
\usepackage{upgreek}

\usepackage[left]{lineno}

\theoremstyle{thmstyleone}%
\theoremstyle{thmstyletwo}%

\theoremstyle{thmstylethree}%

\begin{document}

\title[Article Title]{3D imaging of the biphoton spatiotemporal wave packet}


\author[1,2,3,4]{\fnm{Yang} \sur{Xue}}\email{xuey998@mail.ustc.edu.cn}

\author[1,2,3,4]{\fnm{Ze-Shan} \sur{He}}\email{zshe@mail.ustc.edu.cn}

\author[1,2,3,4]{\fnm{Hao-Shu} \sur{Tian}}\email{hstian@mail.ustc.edu.cn}

\author[1,2,3,4]{\fnm{Qin-Qin} \sur{Wang}}\email{wgqinqin@ustc.edu.cn}

\author[1,2,3,4]{\fnm{Bin-Tong} \sur{Yin}}\email{yinbt@mail.ustc.edu.cn}

\author[1,2,3,4]{\fnm{Jun} \sur{Zhong}}\email{zhong\_jun@mail.ustc.edu.cn}

\author*[1,2,3,4]{\fnm{Xiao-Ye} \sur{Xu}}\email{xuxiaoye@ustc.edu.cn}

\author*[1,2,3,4]{\fnm{Chuan-Feng} \sur{Li}}\email{cfli@ustc.edu.cn}

\author[1,2,3,4]{\fnm{Guang-Can} \sur{Guo}}\email{gcguo@ustc.edu.cn}

\affil[1]{\orgdiv{Laboratory of Quantum Information}, \orgname{University of Science and Technology of China}, \orgaddress{%\street{No.96 Jinzhai}, 
\city{Hefei}, \postcode{230026}, \state{Anhui}, \country{China}}}

\affil[2]{\orgdiv{Anhui Province Key Laboratory of Quantum Network}, \orgname{University of Science and Technology of China}, \orgaddress{%\street{No.96 Jinzhai}, 
\city{Hefei}, \postcode{230026}, \state{Anhui}, \country{China}}}

\affil[3]{\orgdiv{CAS Center for Excellence in Quantum Information and Quantum Physics}, \orgname{University of Science and Technology of China}, \orgaddress{%\street{No.96 Jinzhai}, 
\city{Hefei}, \postcode{230026}, \state{Anhui}, \country{China}}}

\affil[4]{\orgdiv{Hefei National Laboratory}, \orgname{University of Science and Technology of China}, \orgaddress{%\street{No.96 Jinzhai}, 
\city{Hefei}, \postcode{230088}, \state{Anhui}, \country{China}}}

\abstract{
%Possessing a rich number of controllable degrees of freedom (DoFs) makes the photons one of the most important participants in quantum information. %quantum computation, communication and metrology.% for sharing the correlations or even entanglement. 
%Light can possess a rich spatiotemporal structure controllable at room temperature makes it a critical role in both classical and quantum technologies.
Photons are among the most important carriers of quantum information owing to their rich degrees of freedom\,(DoFs), including various spatiotemporal structures. 
The ability to characterize these DoFs, as well as the hidden correlations among them, directly determines whether they can be exploited for quantum tasks.
%This feature, on the one hand, can significantly increase the encoding capacity per photon  by exploiting the correlation between or among these rich DoFs, however, on the other hand, may be harmful for decreasing the purity of the relevant encoding system induced by uncontrollable correlations.
%Spatiotemporal correlations have been deeply studied for classical light field and are believed to ....
%More critically, rich correlations existing among different DoFs significantly increase the encoding capacity of photons. 
%Previous researches have clearly reveal rich correlations among different DoFs may exist in the classical light field. 
%There have been yet no reports of observing such correlation shared by quantum light field, i.e., among photons. 
%Characterising the correlation comes to be a critical technique to improve the performance of photonic quantum system, 
%
While various methods have been developed for measuring the spatiotemporal structure of classical light fields, owing to the technical challenges posed by weak photon flux, there have so far been no reports of observing such structures in their quantum counterparts, except for a few studies limited to correlations within individual DoFs.
Here, we propose and experimentally demonstrate a self-referenced, high-efficiency, and all-optical method, termed 3D imaging of photonic wave packets, for comprehensive characterization of the spatiotemporal structure of a quantum light field, i.e., the biphoton spatiotemporal wave packet.
Benefiting from this developed method, we successfully observe the spatial--spatial, spectral--spectral, and spatiotemporal correlations of biphotons generated via spontaneous parametric down-conversion, revealing rich local and nonlocal spatiotemporal structure in quantum light fields.
This method will further advance the understanding of the dynamics in nonlinear quantum optics and expand the potential of photons for applications in quantum communication and quantum computing.
}

%%================================%%
%% Sample for structured abstract %%
%%================================%%

% \abstract{\textbf{Purpose:} The abstract serves both as a general introduction to the topic and as a brief, non-technical summary of the main results and their implications. The abstract must not include subheadings (unless expressly permitted in the journal's Instructions to Authors), equations or citations. As a guide the abstract should not exceed 200 words. Most journals do not set a hard limit however authors are advised to check the author instructions for the journal they are submitting to.
% 
% \textbf{Methods:} 
% 
% \textbf{Results:} 
% 
% \textbf{Conclusion:} }

%\keywords{keyword1, Keyword2, Keyword3, Keyword4}



\maketitle

\section{Introduction}\label{sec.intro}
The availability of a rich set of degrees of freedom\,(DoFs)\,\cite{Erhard2020,Forbes2025} that are controllable at room temperature underpins the central role of photons in quantum communication\,\cite{Couteau2023,Hu2023}, quantum computation\,\cite{Flamini2019,Romero2024}, and quantum metrology\,\cite{Pirandola2018}.
%Possessing a rich number of degrees of freedom\,(DoFs)\,\cite{Erhard2020,Forbes2025} controllable at room temperature warrants the photons acting as one of the most important participants in quantum communication\,\cite{Couteau2023,Hu2023}, quantum computation\,\cite{Flamini2019,Romero2024}, and quantum metrology\,\cite{Pirandola2018}.
%\,\cite{White1999,Xiang2008,Sipe2012,Reimer2019,Kewming2020,Rambach2021,Karimi2023,Goel2024,Erhard2020,XiaoMin2020,Franson1989,Martin2017,Anton2001,Erhard2018,Baosen2022,yanbo2024,Defienne2021}. 
On the one hand, correlations among these DoFs constitute a scarce and valuable resource for photonic quantum computation; for example, they enable the generation of so-called hyper entanglement\,\cite{kwiat1997,Pan2019}, which can dramatically increase the dimensionality of the accessible Hilbert space\,\cite{Graham2015,Wang2018}.
%On the one hand, correlations among those DoFs act as scarce resources in constructing photonic quantum computers, for instance, generating the so-called hyper entanglement\,\cite{kwiat1997,Pan2019} 
%\,\cite{Graham2015,Camphausen2021,XiaoMin2022,WaXiLin2022,Sheng2023,Shaw2025}, 
%that can significantly increase the size of the relevant Hilbert space\,\cite{Graham2015,Wang2018}.
%\,\cite{Wang2018,Graffitti2020,Karimi2024prl}.
%
On the other hand, although photons, as flight qubits or qudits, are often regarded as relatively resilient to environmental decoherence compared to mass quantum systems, correlation with the unconcerned DoFs may destroy the quantum coherence in the relevant ones, resulting in a reduced purity of the corresponding quantum state\,\cite{Grice2001,Bennink2010,Silberhorn2018oe,Baghdasaryan2023}.
%\,\cite{Grice2001,Walmsley2013,Baghdasaryan2023,Robert2023,Hutter2020,Anand2017,Anand2016,Koefoed2019,Silberhorn2018oe,Bennink2010,Hanzhengfu2024}.
%
As a consequence, it is vital to characterize the joint complex amplitudes spanning the relevant photonic Hilbert space, together with the correlations hidden inside, when photons are employed as quantum information carriers and their nonclassical features are to be exploited.

Moreover, recent advances in pulsed squeezed-light generation\,\cite{Nehra2022} have made sub-cycle quantum features of electromagnetic radiation experimentally accessible, opening the door to a new class of time-dependent quantum states of light\,\cite{Harris2007,Kizmann2019}. Harnessing such nonclassical states as resources for ultrafast quantum optics and quantum information processing likewise requires techniques capable of resolving the full spatiotemporal structure of quantum light fields.
%Besides, the significant advances in pulsed squeezed-light generation reported recently\,\cite{Nehra2022} make the subcycle quantum features of electromagnetic radiation accessible and promises a new class of time-dependent quantum states of light\,\cite{Harris2007,Kizmann2019}. To promote such nonclassical quantum states as a resource for novel ultrafast in quantum optics and quantum information also needs the technique observing the spatiotemporal structure of a quantum state of light.
% for novel ultrafast
Unlike in the classical scenario, where the optical electric field can be fully characterized and even ultrafast temporal structure can be resolved\,\cite{Trebino2002,Walmsley2009,Tang2022,Howard2025}, the intrinsically weak photon flux of nonclassical light fields precludes the direct application of classical measurement techniques, rendering the observation of correlations hidden in quantum states of light particularly challenging.
%Unlike in the classical scenario, where the optical electric field can be characterized thoroughly and even the ultrafast temporal structure can be caught\,\cite{Trebino2002,Walmsley2009,Tang2022,Howard2025}, the weak flux of photons in nonclassical light field inhibits the applications of techniques available in measuring classical correlation, observing the correlations hidden in the quantum state of light faces challenges. 

%One of the most intriguing and promising types of correlation is the spatiotemporal correlation. 
Existing techniques have already pushed the encoding capacity of individual degrees of freedom to their limits, shifting the challenge toward connecting multiple physical dimensions to further expand the accessible Hilbert space. In contrast to the classical light fields, where spatiotemporal correlations are confined within a single wave packet--often referred to as local correlations--in the quantum regime, these correlations are shared between particles, manifesting as quantum nonlocality\,\cite{Pan2012,Killoran2014,Malik2016}.
%\,\cite{Ambrosio2016,Smith2020,Shen2022,White2025,Shen2024,Zhang2024}.
%
Considering a typical case, the quantum state of biphoton can be written as
\begin{eqnarray}
    \label{eqSPDC2}
    \left | \psi \right\rangle = \int d\mathbf{k}_sd\mathbf{k}_i\psi(\mathbf{k}_s,\mathbf{k}_i)%\hat{a}^\dagger_{\mathbf{k}_s}\hat{a}^\dagger_{\mathbf{k}_i}
    |1\text{:}\mathbf{k}_s\rangle|1\text{:}\mathbf{k}_i\rangle.
\end{eqnarray}
Here $|1\text{:}\boldsymbol{f}\rangle=\hat{a}^\dagger_{\boldsymbol{f}}|0\rangle$ with $\hat{a}^\dagger_{\boldsymbol{f}}$ the creation operator on mode $\boldsymbol{f}$ is the short notation for a state with one photon in the corresponding mode and zero in all other modes\,\cite{fabre2020}. The wave vector $\boldsymbol{f} \in\{\mathbf{k}_s,\mathbf{k}_i\}$ defines the mode with the subscripts $s,i$ labeling the photons, which are habitually and hereafter called signal or idler, respectively.
The complex function $\psi(\mathbf{k}_s,\mathbf{k}_i)$ in terms of modes is known as the biphoton wavefunction. Since the wave vectors depend on both spatial and spectral variables, $\psi$ is a biphoton spatiotemporal wave packet after performing the Fourier transformation to the frequency domain. When the biphoton wavefunction cannot be factorized into a direct product of purely spatial and purely temporal components, the biphoton exhibits spatiotemporal correlations.
Although a number of techniques have been reported for measuring the joint complex amplitude in either the spatial\,\cite{Schneeloch2019,Karimi2024prl,Ohad2025,Karimi2024,Karimi2023,Thekkadath2023} or spectral\,\cite{Resch2018,Jin2025,resch2019,walmsley2022,Hurvitz2024,Smith2020} DoF individually, direct experimental observation of spatiotemporal correlations in biphoton quantum states has not yet been demonstrated.
%While there have been reported some technologies measuring the joint complex amplitude in spatial\,\cite{Schneeloch2019,Karimi2024prl,Ohad2025,Karimi2024,Karimi2023,Thekkadath2023} or spectral\,\cite{Resch2018,Zielnicki2018,Jin2025,Donnell2009,resch2019,walmsley2022,Hurvitz2024,Smith2020} DoF individually, observing the spatiotemporal correlation in a biphoton quantum state still lacks presentation. \par 
%
Here, by developing a technique termed joint spatiotemporal amplitude\,(JSTA) measurement, we enable three-dimensional imaging of quantum light fields. As a demonstration, we comprehensively reconstruct the spatiotemporal structure of a typical quantum state of light, namely twin photons generated via spontaneous parametric down-conversion\,(SPDC).
% What do we mean by 3D? The 2D transvers spatial dimension and 1D temporal(spectral) dimension, do not distinguish between signal and idler photons.  
%
First, through cross-phase modulation in a photonic crystal fiber\,(PCF)\,\cite{Matsuda2016}, a constant spectral translation is applied to either the signal or the idler photons while leaving the other unmodified. These two configurations are implemented in separate measurements, enabling full characterization of joint spectral amplitude\,(JSA) using spectral shearing interferometry\,(SSI)\,\cite{Smith2018prl,Smith2020}.
%Firstly, by sequentially and individually introducing a constant spectral translation to the signal or idler photons via the cross-phase modulation in a photonic crystal fiber\,(PCF)\,\cite{Matsuda2016}, the joint spectral amplitude\,(JSA) of biphoton is measured using the technique of spectral shearing interferometry\,(SSI)\,\cite{Smith2018prl,Smith2020}. %\,\cite{Smith2018pra,Smith2018prl,Smith2020}.
% Should we point out the necessity of exchanging the path of signal and idler, and repeat the experiment?
%
Further, if spatial post-selection is introduced prior to the temporal measurement, the spatially resolved spectral amplitude can be obtained. Combined with the spectrally resolved wavefront data, this enables a complete characterization of the spatial--spectral, also the spatiotemporal structure of the biphoton wavefunction. 
This is a self-referenced, highly efficient, and all-optical method which enables joint spatial--spectral measurements at the single-photon level. In the spectral domain, it overcomes the need for an external, well-characterized reference pulse and avoids the low sampling efficiency associated with sum-frequency generation. While in the spatial domain, it preserves the spatial dependence of the temporal envelope, thereby enabling investigation of the joint spatial--spectral structure of the quantum light field.
Owing to this novel technique, we comprehensively characterize the signal photons while post-selecting idler photons in different time-frequency modes and spatial positions. As a result, we successfully observe the spatial--spatial, spectral--spectral, and even spatiotemporal correlations in the biphoton state. To the best of our knowledge, this is the first instance of joint spatial--spectral measurement of biphotons.

\section{Biphoton with spatiotemporal correlation}\label{sec.biphoton}

\begin{figure*}
    \centering
    \includegraphics[width=0.95\linewidth]{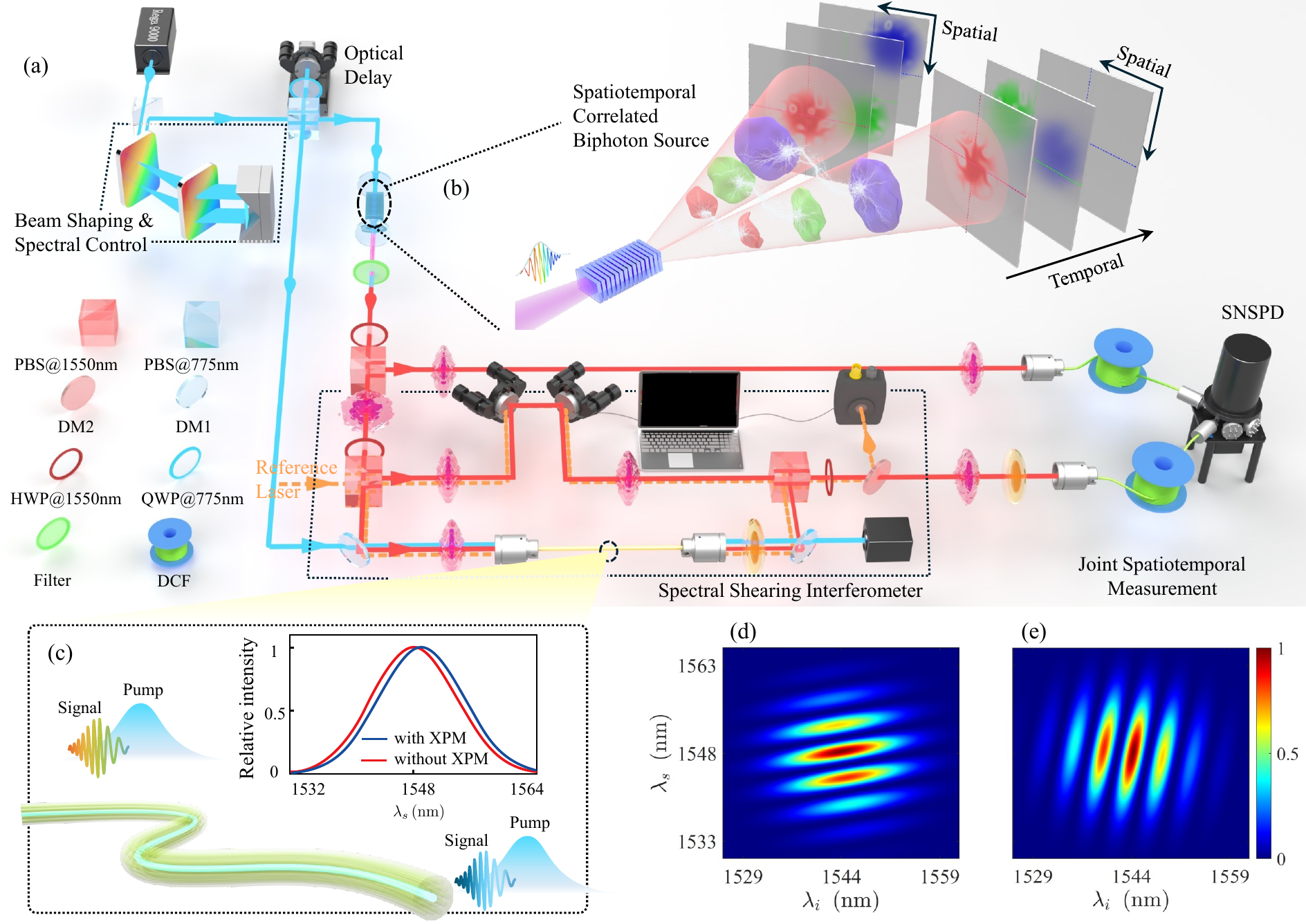}
    \caption{\textbf{Experimental setup and schematic diagram of biphoton spatiotemporal correlations.} As shown in (a), a laser pulse from Rega 9000 is firstly spatially filtered and spectrally dispersed by the beam shaping and spectral control module, then split by a PBS into two beams, one for pumping a cpKTP(custom-poled KTP) crystal to generate the spatiotemporally correlated photon pairs as shown in inset (b), another for pumping the PCF to implement the spectral shearing as shown in inset (c). The biphoton generated in type-\uppercase\expandafter{\romannumeral2} SPDC are in collinear configuration but have perpendicular polarization to each other, as shown by the two red cones in (b). The spatiotemporal correlation is sketched as their correlated spatial intensity distribution varying with the different time, as shown by the different color spots representing the frequencies on the gray temporal slices. To observe these correlations, one photon is directly collected by a fiber coupler and guided to the SNSPD after passing through a 4-km long DCF, while the other passes through the spectral shearing interferometer\,(SSI), as shown by the bottom dashed box in (a). Within the SSI, the photon's amplitude is split into two portions, one undergoes a 25-cm long PCF and, in the presence of the pump pulse, its spectrum is shifted by $\Omega$ due to the nonlinear cross-phase modulation(XPM), the other just propagates in free space with an adjustable temporal delay $\tau$. The two portions of amplitude are coherently recombined before entering the fiber coupler and subsequent DCF and SNSPD. The concept of the spectral shearing process is illustrated in (c), with the inset diagramming the photon's original (red) and shifted (blue) spectrum. At the detection stage, the spatial imaging is implemented by scanning the fiber couplers and, the two DCFs map the photons' frequencies to their arrival times at the detectors which are captured by the coincidence counts with picoseconds resolution. The joint spatial--spectral measurement is finally implemented. (d) and (e) diagram the joint spectral interference patterns of signal (d) or idler (e) photons, respectively, after passing through SSI. The tilted interference fringes indicate the existence of joint spectral phase correlations.}
    \label{fig:1}
\end{figure*}

As a direct consequence of energy and momentum conservation, twin photons generated via SPDC intrinsically exhibit correlations in both space and time, as illustrated in Fig.\,\ref{fig:1}(b), and generally share a global spatiotemporal wavefunction $\psi(\mathbf{k}_s,\mathbf{k}_i)$ as given in Eq.\,\ref{eqSPDC2}.
Following the conventional theoretical treatment\,\cite{Kolobov1999,Howell2016}, the exact form of the normalized global biphoton wavefunction can be obtained from solving the relevant Schr\"odinger equation by utilizing the first-order time-dependent perturbation theory and only taking into account the first order term.  
As shown in the Supplementary Information, $\psi(\mathbf{k}_s,\mathbf{k}_i)$ generally cannot be factorized into a direct product of two wavefunctions of individual modes and depends on both the spatiotemporal profile of the pump pulse and the structure of the interaction region within the crystal. 
It is this nondecomposable feature that requires the nonlinear crystal's structure to be specifically designed to generate entangled photons with high purity\,\cite{Grice2001}.
%\,\cite{Dosseva2016,Cui2019,Graffitti2017,Xu2021,Rozenberg2022,Yesharim2023,Sultanov2024}.
%
Here, we clarify the distinction between the biphoton spatiotemporal wavefunction and the biphoton spatial--spectral wavefunction. Both provide a complete description of the quantum light field and are related to each other by Fourier transformation, differing only in the physical domain in which the observation is performed.

Let us consider a typical experimental scenario in which the transverse dimensions of the nonlinear crystal are much larger than the pump wavelength, and the interaction time, determined by the crystal's longitudinal length together with the chromatic dispersion between the pump and the biphoton, is supposed to be much longer than the temporal duration of the pump pulse. 
These conditions ensure that transverse momentum conservation and energy conservation are exactly satisfied simultaneously.
Thus, the biphoton spatiotemporal wavefunction can then be further simplified as
\begin{equation}
\label{eq:psi2}
    \psi( \mathbf{k}_s,\mathbf{k}_i) = \mathcal{N} \Phi ( \Delta\mathbf{q}, \Delta\nu)\tilde{E}(\mathbf{q}_p,\nu_p).
\end{equation}
$\mathcal{N}$ is a normalization constant. $\Phi$ corresponds to the spatial--spectral phase-matching function, depending on the length and internal structure of the nonlinear crystal. $\tilde{E}$ denotes the spatial--spectral envelope of the pump pulse.
Note that the transverse momentum mismatch $\Delta\mathbf{q}\equiv\mathbf{q}_s-\mathbf{q}_i$ with $\mathbf{q}_s$ and $\mathbf{q}_i$ the photons' transverse momentum component, and the corresponding frequency (energy) mismatch $\Delta\nu\equiv\nu_s-\nu_i$ with $\nu_s,\nu_i$ the relative frequencies of the signal and idler photons.
Note that the spatial--spectral phase-matching function $\Phi$ imposes constraints along the difference directions in transverse spatial momentum and spectral frequency, arising from longitudinal momentum conservation and group velocity matching. 
In contrast, the pump light field $\tilde{E}$ determines the allowed bandwidth along the orthogonal direction, corresponding to the sum of transverse spatial momentum and spectral frequency, as dictated by transverse momentum conservation $\mathbf{q}_p=\mathbf{q}_s+\mathbf{q}_i$ and energy conservation $\nu_p=\nu_s+\nu_i$.
Therefore, the spatiotemporal correlations of the biphoton wavefunction are shaped not only by the properties of the nonlinear crystal but also by the spatial--spectral structure of the pump pulse, a factor that has often been overlooked in previous studies.

Generally, once the nonlinear crystal has been designed and fabricated, the properties of biphoton joint spatiotemporal correlations are primarily influenced by the spatial--spectral envelope of the pump pulse, $\tilde{E} ( {\mathbf{q}}_p,\nu_p )$.
Although spatiotemporally structured pump pulses can induce rich and diverse joint correlations, for simplicity and clarity, we confine our discussion to cases where the spatial and temporal domains are decoupled, leaving the exploration of more complex pump fields to future work. 
In the spatial domain, we assume the pump pulse has a Gaussian transverse intensity profile with a uniform spatial phase distribution, while in the spectral domain, the pump pulse is modeled with a Gaussian spectrum, whose bandwidth is matched to the crystal’s phase-matching function, and exhibits strong dispersion. Under these assumptions, the biphoton wavefunction can be further simplified, as illustrated schematically in Fig.\,\ref{fig:1}(b). The spatial positions of the signal and idler photons exhibit anti-correlation in the transverse momentum domain, while their temporal positions demonstrate positive correlation. Furthermore, both photons inherit the spectral dispersion of the pump pulse, such that their instantaneous frequencies vary over time. Additionally, the individual wave packets of the signal and idler photons exhibit intrinsic spatiotemporal correlations. A detailed description is given in the Supplementary Information.

\section{Joint spatiotemporal amplitude measurement}\label{sec.JSTA}
To observe the aforementioned spatiotemporal correlations, it is necessary to characterize the biphoton joint spatiotemporal amplitude, which requires not only their intensity but also the phase information. %While intensity provides meaningful information at individual positions, only the relative phase between different positions carries physical significance. Consequently, the main challenge lies in measuring the joint spatiotemporal phase.
The latter, i.e., measuring the biphoton's joint spatiotemporal phase faces challenge for the lack of a well-defined phase reference, and the difficulty of performing phase measurements in one dimension while preserving its correlations with the remaining dimensions. 
%the ultrashort temporal duration on the femtosecond scale, the extremely weak electric field strength at the single-photon level, and the difficulty of performing phase measurements along one dimension while preserving correlations with the other dimensions.
%Over \textcolor{red}{the past few decades}, there have developed several methods for measuring the global function of biphotons\,\cite{Altepeter2005}. 
Early studies primarily focused on the joint intensity of biphotons in either the spatial or spectral domain, without considering the corresponding joint phase information\,\cite{Silberhorn2009,Resch2018,Jin2025,Schneeloch2019,Karimi2024prl,Ohad2025}.
Recently developed methods, including non-interferometric ones based on intensity measurements at multiple planes combined with transformation relations between them\,\cite{resch2019,Karimi2024}, and interferometric ones that introduce an additional phase reference\,\cite{walmsley2022,Karimi2023}, have paid attention to measure the joint complex amplitude that can directly give the biphoton wavefunction.
While the non-interferometric approaches often suffer from limited sampling efficiency and resolution at multiple planes, and challenges in extending to the high-dimensional Hilbert space, interferometric techniques can directly extract the phase difference between the unknown state and the reference along the chosen DoFs. When the reference is externally introduced, for example, a classical counterpart attenuated to the single-photon level\,\cite{walmsley2022,Thekkadath2023}, it should be well characterized prior to interference with the unknown state. However, in some cases\,\cite{Karimi2023,Hurvitz2024}, such reference is a resemble of the unknown state, causing the characterization of the reference to become logically circular. Although these works claim to measure the phase information of the unknown state, actually, they merely capture the relative phase variations before and after the artificial manipulation, rather than the initial phase distributions. 
Shearing interferometer is proposed as a self-referenced alternative approach for retrieving the initial phase information\,\cite{Kocsis2011,Smith2020}. The unknown state is translated by a small amount in the spectral or spatial domain and serves as its own reference. Interference between two displaced replicas of the same state reveals the initial phase gradient along the translation direction. By integrating this gradient, the initial phase information can be reconstructed.
However, it should be noted that all of the aforementioned methods are restricted to independent measurements of either the spatial or temporal degrees of freedom, leaving the joint spatiotemporal characterization of biphoton states largely unexplored.

Following the conceptual framework used for classical wave packet characterization, we decompose the joint spatiotemporal phase into two components: the spectrally resolved joint spatial phase $\phi_{\omega_s ,\omega_i}(\mathbf{q}_s, \mathbf{q}_i)$ and the spatially resolved joint spectral phase $\phi_{\mathbf{q}_s, \mathbf{q}_i}(\omega_s ,\omega_i)$, where $\omega_s$ and $\omega_i$ denote the frequencies of signal and idler photons, respectively. The former describes phase variations across the transverse spatial dimensions at specific frequencies, while the latter accounts for phase variations along the longitudinal spectral (or, equivalently, temporal) dimension at specific spatial positions. For femtosecond pulses, measurements are typically performed in the spectral domain due to the limited time resolution of detectors; therefore, we express the joint spatiotemporal phase in the frequency domain in the following. Accordingly, the relative phase between two arbitrary joint spatial--spectral positions $(\mathbf{q}_s,\mathbf{q}_i;\omega_s,\omega_i)$ and $(\mathbf{q}_s',\mathbf{q}_i';\omega_s',\omega_i')$  can be expressed as 
\begin{eqnarray}
\label{eq:JSSP}
&&\Delta \phi\left[(\mathbf{q}_s',\mathbf{q}_i';\omega_s',\omega_i') \rightarrow (\mathbf{q}_s,\mathbf{q}_i;\omega_s,\omega_i)\right] \notag\\
&=& \Delta \phi_{\mathbf{q}_s, \mathbf{q}_i}(\omega_s \to \omega_s' ,\omega_i) 
+ \Delta \phi_{\omega_s', \omega_i}(\mathbf{q}_s \to \mathbf{q}_s',\mathbf{q}_i)\notag
\\ 
 &+& \Delta \phi_{\mathbf{q}_s', \mathbf{q}_i}(\omega_s' ,\omega_i \to \omega_i') 
+ \Delta \phi_{\omega_s', \omega_i'}(\mathbf{q}_s',\mathbf{q}_i \to \mathbf{q}_i').
\end{eqnarray}
Here, the subscripts appearing in the equation's right side stands for that the relative phase is given at the corresponding position or frequency. 
%$\triangle$ denote the phase changes from positions on the left side of $\to$ to positions on the right side of $\to$. 
For instance, $\Delta \phi_{\mathbf{q}_s, \mathbf{q}_i}(\omega_s \rightarrow \omega_s' ,\omega_i) \equiv \phi_{\mathbf{q}_s, \mathbf{q}_i}(\omega_s' ,\omega_i) -\phi_{\mathbf{q}_s, \mathbf{q}_i}(\omega_s,\omega_i)$ denotes the relative phase between the signal frequency components $\omega_s'$ and $\omega_s$, evaluated at a fixed transverse position $\mathbf{q}_s$, with the idler photon conditioned at the spatial–spectral coordinate $(\mathbf{q}_i,\omega_i)$.
The first and third terms in Eq.\,\ref{eq:JSSP} could be obtained via the spatially resolved joint spectral amplitude measurements, while the second and fourth terms could be obtained via the spectrally resolved joint spatial amplitude measurements. By combining these two procedures, the joint spatial--spectral phase of the biphoton state could be fully reconstructed. 

\begin{figure*}
    \centering
    \includegraphics[width=0.8\textwidth]{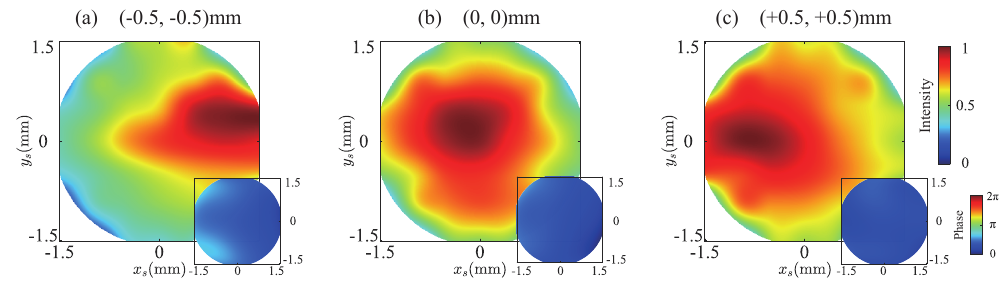}
    \caption{ \textbf{Biphoton joint spatial correlations.} The spatial intensity distribution of the signal photons while the idler photon are post-selected at (a) (-0.5, -0.5)\,mm, (b) (0, 0)\,mm and (c) (+0.5, +0.5)\,mm three specific spatial positions. The spectral dimension is neglected for clarity. The inset in the lower-right corner illustrate the corresponding spatial phase distribution.}
    \label{fig:2}
\end{figure*}

\begin{figure*}
    \centering
    \includegraphics[width=1\linewidth]{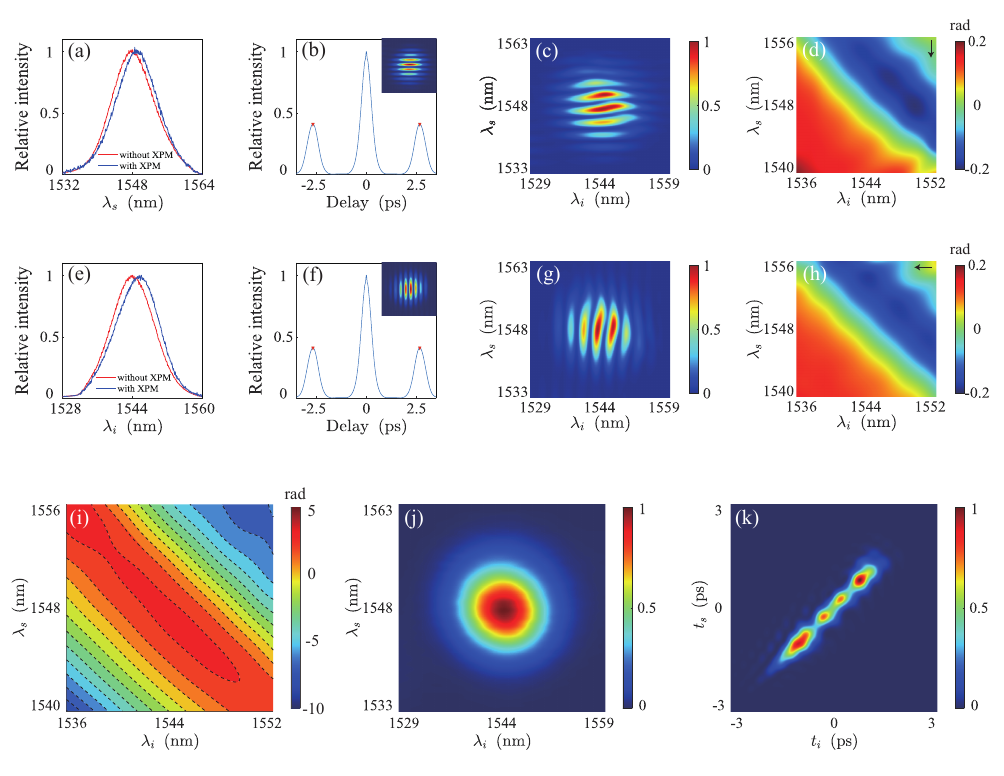}
    \caption{ \textbf{Biphoton joint spectral correlations.} Measurement process of the joint spectral phase gradient along the signal axis(a-d) and idler axis(e-h) when both of them are post-selected in the central spatial positions. The reconstructed joint spectral and temporal amplitude are illustrated in (i-k).  (a) and (e) show the spectra of the signal and idler photons before(red) and after(blue) spectral shearing. (b) and (f) display the Fourier transformation of the joint spectral interference patterns without spectral shearing(shown in the insets), along the signal and idler axes, respectively. The peak location of the side-lobes denote the temporal delay between the two arms of SSI. (c) and (g) show joint spectral interference patterns with spectral shearing of signal and idler photons. with calibrated frequency shift and temporal delay, the joint spectral phase gradients along the signal and idler axes are quantitatively determined, as shown in (d) and (h), respectively. The arrows indicate the gradient direction. (i) The joint spectral phase distribution reconstructed from the two-dimensional gradient field composed of (d) and (h). (j) The directly measured joint spectral intensity. (k) The joint temporal intensity obtained by Fourier transforming the joint spectral complex amplitude composed of (i) and (j) from the spectral domain to the temporal domain. }
    \label{fig:3}
\end{figure*}

Here, we extend the spectral shearing interferometer\,\cite{Smith2020} to the spatial domain by replacing the collinear interferometer with a Mach–Zehnder interferometer configuration, enabling joint spatiotemporal measurement, as shown in Fig.\,\ref{fig:1}(a). In our scheme, the signal photons propagate through the SSI to characterize their local spatial--spectral complex amplitude, conditioned on coincident idler photons post-selected in different spectral modes and spatial positions. 
The key role within the SSI is the photonic crystal fiber. To perform the spatially resolved joint spectral phase measurements, a portion of the signal photons copropagates with a classical control pulse and is temporally aligned with its rising edge, as shown in Fig.\,\ref{fig:1}(c). This results in an approximately linear cross-phase modulation in the temporal domain, giving rise to a constant frequency shift in the spectral domain\,\cite{Matsuda2016}, as illustrated in the inset. The sheared portion serves as a phase reference, enabling self-referenced interference between the original and frequency-shifted states. The resulting interference pattern encodes the spectral phase gradient, as shown in Fig.\,\ref{fig:1}(d), thereby determining the phase variation along the signal axis and leaving phase variations along the orthogonal axis undetermined. This could be complemented by exchanging the roles of the signal and idler photons and repeating the measurement, whereby the corresponding phase gradient along the idler axis could be extracted from the interference pattern shown in Fig.\,\ref{fig:1}(e). Combining these two orthogonal gradients together, the spatially resolved joint spectral phase could be retrieved by zonal approach\,\cite{Southwell1980}.
Meanwhile, to perform spectrally resolved joint spatial phase measurements, the photonic crystal fiber also acts as a fundamental Gaussian spatial mode filter. When the classical control pulse is blocked, the portion of signal photons transmitted through the PCF provides a well-defined spatial phase reference for the original beam. Due to the sensitivity of the interference fringes to the photon arrival time, when the fiber coupler scans through the entire space, the changes of the position of the interference fringes encode the spatial phase distribution, enabling the retrieval of the spectrally resolved joint spatial phase. (See Methods and the Supplementary Information for more details)

\section{Experimental Results}\label{sec.results}
As mentioned above, the full characterization of the joint spatiotemporal correlations consists of two parts: the spectrally resolved joint spatial measurement and the spatially resolved joint spectral measurement. We first measure the complex spectrally resolved joint spatial amplitude, including both intensity and phase. The signal spatial intensity is measured with the PCF arm of SSI completely blocked, while the spatial phase is retrieved using the portion of signal photons transmitted through the PCF as a spatial phase reference; in this case, the classical control pulse remains blocked. The results are illustrated in Fig.\,\ref{fig:2}. For clarity, since the primary observational perspective at this stage lies in the spatial domain, we integrate over the spectrum to obtain the spatial intensity distribution and average the spectral phase to extract the wavefront distribution.
The measured joint spatial intensity exhibits a clear anti-correlation between the signal and idler photons: the geometrical center of the signal photons’ spatial intensity distribution shifts in the direction opposite to that of the post-selected spatial position of the idler photons. This behavior originates from transverse momentum conservation in the SPDC process. Small differences in displacement along the $x$ and $y$ axes may arise from distortions of the biphoton wave packet during free-space propagation. At the same time, the spatial phase distribution of the signal photons remains nearly uniform, independent of the post-selected spatial position of the idler photons. This absence of spatial phase correlation reflects the uniform wavefront of the pump pulse.

It is worth noting that such joint spatial correlations are often overlooked in bulk-optical systems. Although some studies have investigated the influence of pump spatial parameters in the SPDC process\,\cite{Hutter2020,Bennink2010}, 
%\,\cite{Robert2023,Hutter2020,Bennink2010,Bornman2021,Coccia2023}
attention to these correlations is typically paid only when the spatial DoF is employed as an information carrier\,\cite{OhadLib2024,Brandt2020}. Our experimental results once again emphasize the critical roles of the waist at the focal plane and the spatial wavefront distribution of the pump pulse.
In addition, although some previous studies have reported orbital angular momentum correlations between signal and idler photons, our measurements are performed directly in the Cartesian coordinate system. The apparent discrepancy may arise from differences in the measurement basis. These two results are not mutually exclusive.

%\begin{comment}
%\begin{figure*}
%    \centering
%    \includegraphics[width=0.8\textwidth]{Fig Joint Spatial Amplitude v1.5 c.pdf}
%    \caption{ \textbf{Biphoton joint spatial correlations.} The spatial intensity distribution of the signal photons while the idler photon are post-selected at (a) (-0.5, -0.5)mm, (b) (0, 0)mm and (c) (+0.5, +0.5)mm three specific spatial positions. The spectral dimension is neglected for clarity. The inset in the lower-right corner illustrate the corresponding spatial phase distribution.}
%    \label{fig:2}
%\end{figure*}
%\end{comment}

%\emph{Illustrate the joint spectral phase correlations of biphoton. The joint spectral intensity is uncorrelated, while the joint spectral phase is correlated.}
%
As another part of the biphoton state characterization, spatially resolved joint spectral measurements is performed when the photons passing through the PCF experience cross-phase modulation and serve as a spectral phase reference. The measurement process and results are illustrated in Fig.\,\ref{fig:3}. Take the signal photons passing through the SSI as an example, the corresponding joint spectral interference pattern is shown in Fig.\,\ref{fig:3}(c). The tilted interference fringes qualitatively indicate the correlations in the joint spectral phase. Compared to the simulated results in Fig.\,\ref{fig:1}(d), the noticeable curvature of these fringes reveals the existence of higher-order joint spectral phase correlations. Similar curvature is also observed in Fig.\,\ref{fig:3}(g), which is a strong evidence that these correlations originate from the higher-order spectral dispersion of the pump pulse. One potential origin is the third-order dispersion introduced by the grating pairs during temporal stretching in the spectral control module of Fig.\,\ref{fig:1}(a).
With the frequency shift and temporal delay accurately calibrated, the spectral phase gradients along the signal and idler axes could be quantitatively extracted from the joint spectral interference patterns, as shown in Fig.\,\ref{fig:3}(d) and (h). Each gradient individually determines phase variation along one axis while leaving the variation along the orthogonal axis undetermined; this missing information could be complemented by the gradient measured along the other axis. Combining both datasets, the joint spectral phase could be reconstructed from this two-dimensional phase gradient field, as shown in Fig.\,\ref{fig:3}(i). 
Fitting the reconstructed phase yields an estimate of the pump pulse Group Delay Dispersion\,(GDD) of $\mathrm{GDD}_\text{exp}  = -2.66\times 10^5$\,fs$^2$, which agrees well with the theoretical value calculated from the grating pairs parameters $\mathrm{GDD}_\text{th}  = -2.59\times 10^5 $\,fs$^2$.

These results emphasize the critical role of the pump's spectral dispersion. The joint spectral amplitude is shaped not only by the phase-matching function, but also by the spectral properties of the pump pulse. While the cpKTP crystal is engineered to produce an uncorrelated joint spectral intensity, as shown in Fig.\,\ref{fig:3}(j), the spectral dispersion of the pump pulse induces significant correlations in the joint spectral phase, as shown in Fig.\,\ref{fig:3}(i). When transformed into the temporal domain, it manifests as strong joint temporal intensity correlations, as shown in Fig.\,\ref{fig:3}(k). Such correlations cannot be detected through JSI measurements alone, underscoring that intensity-only characterization is insufficient for SPDC sources, which only focus on the pump pulse spectral intensity bandwidth matching that of the phase-matching function, neglecting the role of the pump pulse spectral phase. In other words, a complete characterization of the pump pulse---including both its spectral intensity and phase---is essential prior to nonlinear interaction.

\begin{figure}
    \centering
    \includegraphics[width=0.47\textwidth]{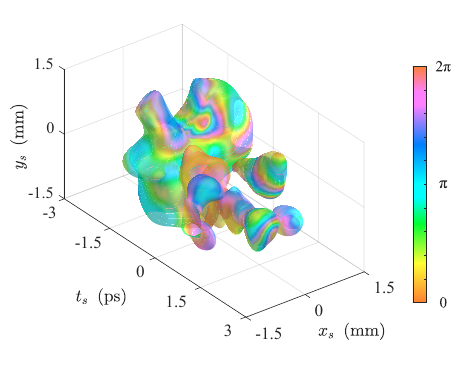}
    \caption{ \textbf{Single photon spatiotemporal correlations.} The spatiotemporal iso-intensity envelope of the signal photons with idler photons post-selected at the central spatial position. Colors represent the corresponding phase distributions along the iso-intensity contour.}
    \label{fig:4}
\end{figure}

\begin{figure}
    \centering
    \includegraphics[width=0.5\textwidth]{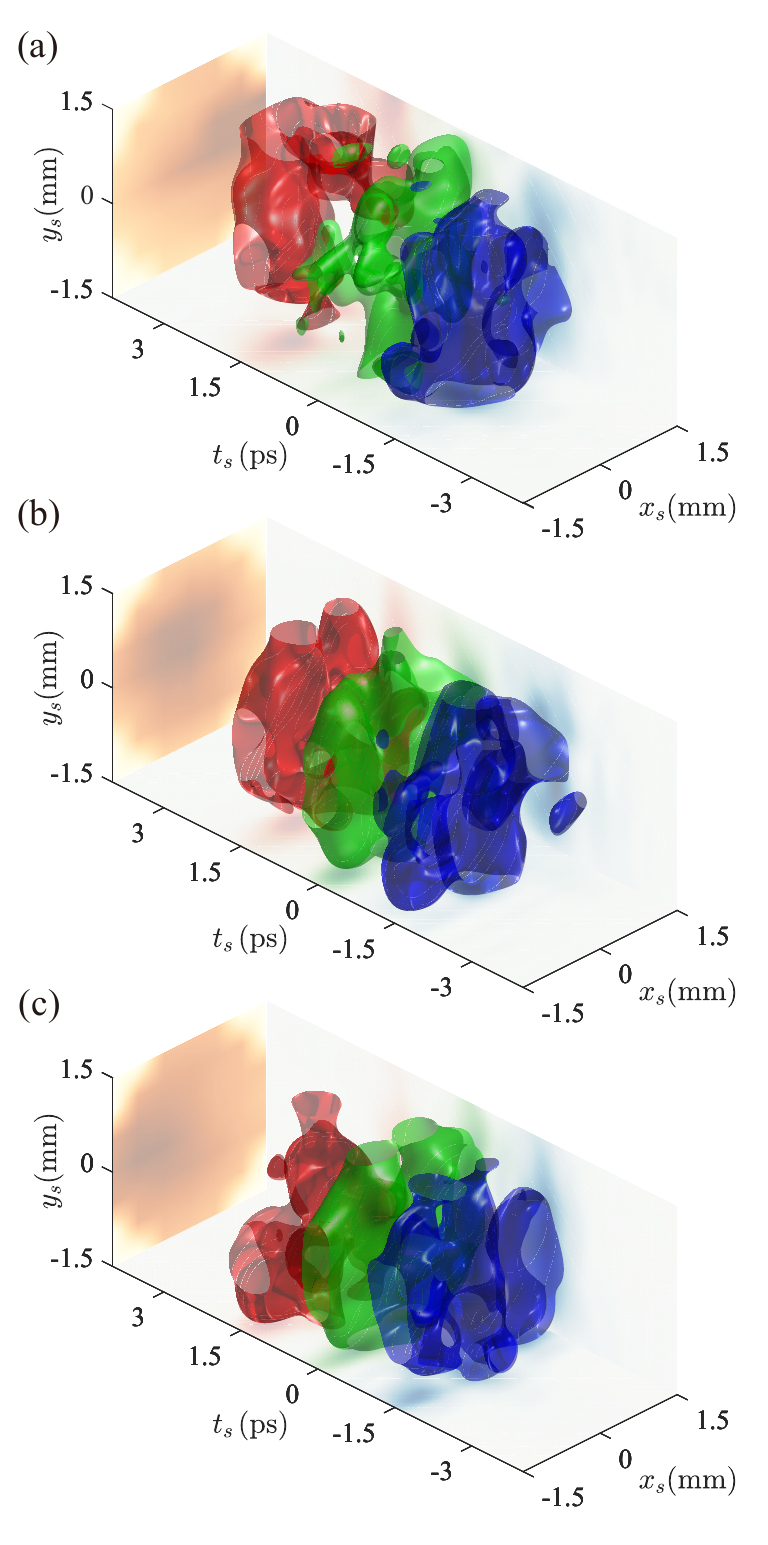} 
    \caption{ \textbf{Biphoton spatiotemporal correlations.} The iso-intensity envelopes of signal photons are shown for idler photons post-selected at spatial positions of (a) (-0.5, -0.5)\,mm, (b) (0, 0)\,mm, and(c) (+0.5, +0.5)\,mm, and temporal positions of -1.5\,ps(red), 0\,ps(green) and +1.5\,ps(blue). Each iso-intensity envelope is normalized to its maximum value and projected onto the $x-t$ and $y-t$ planes. Additionally, the total spatial intensity distribution of the signal photons is illustrated on the $x-y$ plane. }
    \label{fig:5}
\end{figure}

At this stage, combined the results of spectrally resolved joint spatial measurements and spatially resolved joint spectral measurements together, the complex joint spatiotemporal amplitude of the biphoton state is reconstructed. We first illustrate the spatiotemporal amplitude of the signal photons in Fig.\,\ref{fig:4}, and then show how this wave packet varies with the change of the spatiotemporal post-selection of the idler photons in Fig.\,\ref{fig:5}. In Fig.\,\ref{fig:4}, the temporal profiles of signal photons vary with their spatial positions. We term this kind of correlations within single wave packet as `local' spatiotemporal correlations. There are two primary factors contributing to this local spatiotemporal correlations. First, the transverse wave vector $\mathbf{q}$ and the frequency $\omega$ are intrinsically coupled in the phase-matching function, which cannot be factorized into a direct product of separate spatial and spectral components. Second, the signal wave packet may experience distortions during free-space propagation, such as optical aberrations of the lenses, angular dispersion introduced by prisms or gratings, and other system imperfections. These effects further complicate the spatiotemporal structure of signal photons. Nevertheless, the main features remain evident. Because the idler photons are post-selected near the spatial origin, the signal wave packet is correspondingly centered in space. Along the temporal axis, the signal wave packet features a dominant central envelope followed by a series of gradually decaying secondary envelopes, which is a characteristic of third-order dispersion inherited from the pump pulse.

In Fig.\,\ref{fig:5}, it is apparent that the spatiotemporal profile of the signal photons varies with the different spatiotemporal post-selections of the idler photons. We term this kind of correlations shared between signal and idler photons as `nonlocal' spatiotemporal correlations. Compared to Fig.\,\ref{fig:4}, where the idler photons are not post-selected in the temporal domain, the signal wave packet in Fig.\,\ref{fig:5} exhibits a significantly reduced temporal width, approaching the Fourier transform limit and demonstrating strong temporal localization. Furthermore, the central temporal position of the signal wave packet aligns closely with that of the post-selected idler photons, confirming strong temporal correlations. In contrast, the spatial correlations appear less regular. This apparent irregularity arises from the intrinsic local spatiotemporal correlations within the signal and idler wave packets. 
Specifically, the strongest portion of joint temporal intensity at a certain joint spatial position may deviate from the original position and become distorted due to the individual spectral dispersion of either signal or idler photons, and this deviation and distortion may depend on spatial positions. However, since the temporal post-selection of the idler photons is fixed to the same time across the entire space, the corresponding temporal slices of the signal in JTI do not necessarily intersect the peak intensity at different spatial positions. This mismatch results in the distortion of the iso-intensity contours of the signal wave packet.
However, when the signal wave packet is projected along the temporal axis onto the spatial domain, the result collapses to the spatial intensity distribution. Under this projection, the previously observed spatial anti-correlations between the signal and idler photons re-emerge, consistent with the results shown in Fig.\,\ref{fig:2}.
These observation results suggest that the biphoton state exhibits a hybrid structure, containing both local and nonlocal spatiotemporal correlations. A relatively simple explanation could be given that the transverse momentum, $\mathbf{q}_s$ and $\mathbf{q}_i$, and the frequency, $\omega_s$ and $\omega_i$, of the signal and idler photons are intrinsically coupled by phase-matching function and spatiotemporal structure of the pump pulse. During the subsequent individual propagation in free space, the local operations of optical elements between the spatial and temporal dimensions further complicate the spatiotemporal structure of biphoton state. A deeper investigation into the influences of various optical elements is therefore left for future work.

%\section{Discussion}\label{sec12}

%Discussions should be brief and focused. In some disciplines use of Discussion or `Conclusion' is interchangeable. It is not mandatory to use both. Some journals prefer a section `Results and Discussion' followed by a section `Conclusion'. Please refer to Journal-level guidance for any specific requirements. 

\section{Conclusion}\label{sec13}
In conclusion, we propose a self-referenced, high-efficiency, and all-optical approach for measuring the joint spatiotemporal amplitude of biphoton wave packets. To the best of our knowledge, this work represents the first demonstration of three-dimensional imaging of a quantum light field.
%In conclusion, we propose an original, reference-free method for measuring the joint spectral complex amplitude, demonstrating the powerful capability of PCFs in manipulating photon spectra. We extend this method to the spatial domain, marking, to the best of our knowledge, the first instance of measuring the joint spatial-spectral complex amplitude.
Using this approach, we observe both local spatiotemporal correlations within a single photonic wave packet and nonlocal spatial--spatial, spectral--spectral, and spatiotemporal correlations shared between the signal and idler photons. 
%Through this approach, we observe a variety of rich correlations between signal and idler photons, including spatial-spatial, spectral-spectral, and spatial-spectral correlations. 
These observations emphasize the critical role played not only by the nonlinear crystal but also by the pump pulse in the SPDC process, which is often overlooked in previous studies.
Nevertheless, a more quantitative understanding of the role played by the pump pulse’s spatial--spectral structure in shaping biphoton states remains an open question.
Our approach expands the toolkit for characterizing quantum light fields. It is expected to enable access to higher-dimensional photonic Hilbert spaces, thereby facilitating the use of photons as information carriers in quantum communication and quantum computation.
Moreover, our approach is directly applicable to the high-gain regime\,\cite{Triginer2020}, where it may provide deeper insight into the dynamics of quantum nonlinear optics\,\cite{Chang2014}. 
%

%We anticipate that this approach will be applicable to more complex structured photonic wave packets, and this will be the focus of our future work. Our results may help investigate the SPDC process in high gain regime\,\cite{Quesada2014,Quesada2022,Triginer2020,Quesada2020,Amooei2025,Sharapova2020}, and prompt the research of a new class of time-dependent quantum states of light\,\cite{Harris2007,Kizmann2019}. 

\backmatter

\bmhead{Supplementary information}

Supplementary information is available for this paper at xxx.

\bmhead{Acknowledgements}
The authors acknowledge insightful comments from Andrew White. 
This work was supported by Quantum Science and Technology-National Science and Technology Major Project (No.\,2021ZD0301200), National Natural Science Foundation of China (Nos.\,12474494, 12204468), Fundamental Research Funds for the Central Universities (No.\,WK2030000081), and China Postdoctoral Science Foundation (No.\,2024M753083, BX20240353).

\bibliography{references}% common bib file

\section{Methods}\label{sec11}

\textbf{Detailed Experimental setup}\\
\noindent
The pump pulse derived from Rega 9000A centers at 773\,nm, with spectrum bandwidth 5.1\,nm and repetition frequency 250\,kHz, is spatially filtered by two lenses and an pinhole into fundamental Laguerre–Gaussian mode with a waist of 2.5\,mm, and then temporally extended to $\sim$3\,ps by a pair of gratings\,(Gitterwerk, 1379.3\,l/mm\,@\,800\,nm). This pump pulse is weakly focused onto a 5-mm-long cpKTP crystal with a waist of 250\,$\upmu$m with average power 50\,mW. The spatiotemporally correlated biphoton state is generated through type-\uppercase\expandafter{\romannumeral2} SPDC process, with the central wavelength of signal\,(H) and idler\,(V) photons 1548\,nm and 1544\,nm, respectively. The HWP and PBS following the cpKTP crystal controls the paths of photons by their polarization. In the reflected path of the PBS, the fiber coupler\,(Thorlabs, F260FC-1550) changes between three positions: (-0.5, -0.5)\,mm, (0, 0)\,mm, (+0.5, +0.5)\,mm to perform spatial post-selection. The following DCF\,(YOFC, AD-SM-C-040-FC/PC-27/N-02) introduces a frequency-dependent delay of approximately 667\,ps/nm with a transmission efficiency of $\sim$50\%. Considering the timing jitter of the single-photon counting module\,(PHOTEC, SPOT-SYS-10) to be 110\,ps and the time-to-digital converter\,(PicoQuant, HydrapHarp 400) to be 25\,ps, the resulting spectral resolution is 0.169\,nm. In the transmitted path of the PBS, a classical control pulse with average power of 5\,mW within the PCF\,(Photonics Bretagne, SUP-5-125) is used to translate the spectrum of photons by $\sim$-150\,GHz. A temporal delay of $\sim$2.5\,ps is introduced between the two arms of SSI. Every time before measuring the spatially resolved joint spectral interference pattern, the frequency shift and temporal delay are carefully calibrated. Since the interference pattern is phase-sensitive, a continuous laser centered at 1590\,nm\,(Precilasers, FL-SF-795-1-CW) is used for phase-locking. This phase-locking laser is separated from the signal photons by a grating\,(Gitterwerk, 1000\,l/mm\,@\,1560\,nm)). The fiber coupler\,(Thorlabs, F260FC-1550) at the output of the SSI scans the entire space at $7 \times 7$ positions at an interval of 0.5\,mm. The following DCF\,(YOFC, AD-SM-C-080-FC/PC-27/N-02) introduces a frequency-dependent delay of approximately 1343\,ps/nm with a transmission efficiency of $\sim$30\%, corresponding to a spectral resolution of 0.084\,nm. The max coincident counts rate is approximate $\sim$1k\,cps when both the signal and idler photons are spatially post-selected at central position, corresponding to a integration time of 30\,min. As the fiber coupler moves to the edge of the spatial distribution, the coincident counts rate reduce gradually and the integration time increases correspondingly, finally arrive at $\sim$110\,cps and 2.5\,h.

\vskip 10pt

\noindent \textbf{Data analysis}\\
\noindent
We reduce the shot noise in the obtained joint spectral interference pattern by performing two-dimensional Fourier transformation and removing the high frequency components. When performing the spatially resolved joint spectral measurement, the joint spectral phase gradient could be extracted from the interference pattern by Fourier transform filtering techniques in \cite{Walmsley2009} and the Supplementary Information. When performing the spectrally resolved joint spatial measurement, we record the positions of the $n$ interference peaks as $(\lambda_{s1},\lambda_{s2},\lambda_{s3}, ..., \lambda_{sn})$. Since these interference fringes translate by one period $\triangle \lambda_{s}$ when the wavefront changes by $2\pi$, the average of the change of these positions, $(\triangle \lambda_{s1}+\triangle \lambda_{s2}+...+\triangle \lambda_{sn})/n$, divided by a interference period, $\triangle \lambda_{s}$, is proportional to the signal spatial phase variation across two-dimensional space. Our sampling rate in the spatial dimension is relatively sparse, for clarity, we perform interpolation of joint spatial--spectral intensity and phase separately in the spatial dimensions, and the recombine them as complex joint spatial--spectral amplitude. Further, we perform Fourier transformation along the spectral axis to transfer the observation perspective from spatial--spectral to spatiotemporal dimensions.


\end{document}

%\preprint{APS/123-QED}

\title{Supplementary Material for \emph{3D imaging of the biphoton spatiotemporal wave packet}}

\author{Yang Xue}
\author{Ze-Shan He}
\author{Hao-Shu Tian}
\author{Qin-Qin Wang}
\author{Bin-Tong Yin}
%\author{Yukuan Zhao}
\author{Jun Zhong}
\author{Xiao-Ye Xu}
\email{xuxiaoye@ustc.edu.cn}
\author{Chuan-Feng Li}
\email{cfli@ustc.edu.cn}
\author{Guang-Can Guo}
\affiliation{Laboratory of Quantum Information, University of Science and Technology of China, Hefei 230026, China}
\affiliation{Anhui Province Key Laboratory of Quantum Network, University of Science and Technology of China, Hefei 230026, China}
\affiliation{CAS Center for Excellence in Quantum Information and Quantum Physics, University of Science and Technology of China, Hefei 230026, China}
\affiliation{Hefei National Laboratory, University of Science and Technology of China, Hefei 230088, China}

\date{\today}

\maketitle

\section{Joint Spatiotemporal Wave Function of Biphoton State}
In this section, we present a detailed derivation of the joint spatiotemporal wave function of the biphoton state. We begin with the classical Hamiltonian of the electromagnetic field \cite{howell2016}:
\begin{align}
\label{SM Classical H}
\mathcal{H}_{EM} = \mathcal{H}_{L} + \mathcal{H}_{NL},
\end{align}
where $\mathcal{H}_{L}$ is the linear part of the Hamiltonian, and $\mathcal{H}_{NL}$ is the nonlinear contribution given by:
\begin{eqnarray}
\label{SM Classical H_NL}
\mathcal{H}_{NL} =&&\frac{1}{2(\sqrt{2\pi})^3}\epsilon_0  \int \mathrm{d}^3r  
\sum_{\vec{\mathbf{k}}_p,\vec{\mathbf{k}}_1,\vec{\mathbf{k}}_2}\notag
\\
&&\left[\tilde{\chi}_{ijl}^{(2)}(\vec{\mathbf{r}};~\omega(\vec{\mathbf{k}}_p),\omega(\vec{\mathbf{k}}_1),\omega(\vec{\mathbf{k}}_2))
~E_i(\vec{\mathbf{r}},\omega(\vec{\mathbf{k}}_p))
E_j(\vec{\mathbf{r}},\omega(\vec{\mathbf{k}}_1))
E_l(\vec{\mathbf{r}},\omega(\vec{\mathbf{k}}_2))\right].
\end{eqnarray}
Here, $\tilde{\chi}_{ijl}^{(2)}$ denotes the nonlinear susceptibility which depends on the frequencies of the pump, signal and idler fields. Note that $\vec{\mathbf{k}}$ denotes the wave vector, $\omega$ denotes the corresponding angular frequency, and the subscripts p, 1 and 2 label the pump, signal and idler modes, respectively. 

Upon quantization of the electromagnetic field, each field is decomposed into its positive- and negative-frequency components, denoted by the superscripts + and -, respectively. The pump field is assumed to be sufficiently intense and is therefore treated as a classical, undepleted field, $E_p(r,t) = E_p^{+}(r,t) + E_p^{-}(r,t)$, while the signal and idler fields, $E_{1,2}(\vec{\mathbf{r}},t)$, are identified as observables $\hat{E}_{1,2}(\vec{\mathbf{r}},t) = \hat{E}_{1,2}^{+}(\vec{\mathbf{r}},t) + \hat{E}_{1,2}^{-}(\vec{\mathbf{r}},t)$, with the positive-frequency part expressed as:  
%
\begin{comment}
\begin{align}
\label{SM Quantization}
\hat{E}^{+}(\vec{\mathbf{r}},t) = \frac{1}{V^{\frac{1}{2}}} \sum_{\vec{\mathbf{k}},\alpha} i\sqrt{\frac{\hbar \omega (\vec{\mathbf{k}})}{2\epsilon_0 } }~\hat{a}_{\vec{\mathbf{k}},\alpha
}(t) ~\vec{\mathbf{\epsilon}}_{\vec{\mathbf{k}},\alpha}~e^{i\vec{\mathbf{k}} \cdot \vec{\mathbf{r}}} 
\end{align}
\end{comment}
%
\begin{align}
\label{SM Quantization}
\hat{E}^{+}(\vec{\mathbf{r}},t) = \frac{1}{V^{\frac{1}{2}}} \sum_{\vec{\mathbf{k}}_,\alpha} i\sqrt{\frac{\hbar \omega (\vec{\mathbf{k}})}{2\epsilon_0 } }~\hat{a}_{\vec{\mathbf{k}},\alpha }(t) ~\vec{\mathbf{\epsilon}}_{\vec{\mathbf{k}},\alpha}~e^{i\vec{\mathbf{k}} \cdot \vec{\mathbf{r}}},
\end{align}
and $\hat{E}^{-}(\vec{\mathbf{r}},t)$ is its Hermitian conjugate. Here, $V$ is the quantization volume, $\alpha$ labels the polarization components, $\hat{a}_{\vec{\mathbf{k}},\alpha}(t)$ is the photon annihilation operator at time $t$ and $\vec{\mathbf{\epsilon}}$ denotes the corresponding unit polarization vector.

Consequently, the nonlinear part of the Hamiltonian can be extended to eight terms. Retaining only those that satisfy energy conservation, it can be written as:
\begin{align}
\label{SM Quantum H1}
\hat{H}_{NL} = & \frac{1}{2}\epsilon_0 \int \mathrm{d}^3r \Bigg(
\frac{-1}{V}
\sum_{\vec{\mathbf{k}}_1,\alpha_1}
\sum_{\vec{\mathbf{k}}_2,\alpha_2}
\tilde{\chi}_{ijl}^{(2)}(\vec{\mathbf{r}};
\omega(\vec{\mathbf{k}}_p),\omega(\vec{\mathbf{k}}_1),\omega(\vec{\mathbf{k}}_2))
\sqrt{\frac{\hbar^{2}\omega(\vec{\mathbf{k}}_1)\omega(\vec{\mathbf{k}}_2)}{4\epsilon_0^{2}}}
\notag
\\ &
\times e^{-i(\vec{\mathbf{k}}_1+\vec{\mathbf{k}}_2)\cdot\vec{\mathbf{r}}}
E_i(\vec{\mathbf{r}},t)
\hat{a}^\dagger_{\vec{\mathbf{k}}_1,\alpha_1}(t)
\hat{a}^\dagger_{\vec{\mathbf{k}}_2,\alpha_2}(t)
(\vec{\mathbf{\epsilon}}_{\vec{\mathbf{k}}_1,\alpha_1})_j
(\vec{\mathbf{\epsilon}}_{\vec{\mathbf{k}}_2,\alpha_2})_l
+ \mathrm{h.c.}
\Bigg)
\end{align}

Here, h.c. denotes the Hermitian conjugate. Meanwhile, the linear part of the Hamiltonian can be expressed as $\hat{H}_{L} = \sum_{\vec{\mathbf{k}},\alpha}^{}\hbar \omega ( \vec{\mathbf{k}} ) \hat{a}_{\vec{\mathbf{k}},\alpha}^{\dagger} \hat{a}_{\vec{\mathbf{k}},\alpha}$.

During the SPDC process, the pump pulse can be treated as a propagation-invariant spatiotemporal wave packet. We define the transverse momentum $\vec{\mathbf{q}}$ as the component of momentum perpendicular to the propagation direction, the longitudinal momentum $\vec{\mathbf{k}}_z$ as the component parallel to the propagation direction, and the relative frequency $\nu = \omega - \Omega$ as the difference between the frequency $\omega$ and central frequency $\Omega$.
With the pump pulse polarization vector $(\vec{\mathbf{\epsilon}}_{{\vec{\mathbf{k}}_p}})_{i}$ fixed, the spatiotemporal profile of the pump field, $ E_i(\vec{\mathbf{r}},t)$, could be expressed by performing a three-dimensional Fourier transformation to $\tilde{E}_i(\vec{\mathbf{q}},\nu_p)$ over transverse momentum and frequency axis:
\begin{align}
\label{SM Pump Pulse}
E_i(\vec{\mathbf{r}},t) = \frac{1}{(\sqrt{2\pi})^3}\int \mathrm{d}^2 q_p \mathrm{d} \nu_p 
\tilde{E}(\vec{\mathbf{q}},\nu_p) (\vec{\mathbf{\epsilon}}_{{\vec{\mathbf{k}}_p}})_{i}
~e^{i (\vec{\mathbf{q}}_p \cdot \vec{\mathbf{r}} - \nu_p t) }
e^{i (k_{zp} z - \Omega_p t) } .
\end{align}

Substituting Eq.\,\ref{SM Pump Pulse} into Eq.\,\ref{SM Quantum H1}, the quantum Hamiltonian could be rewritten as:
\begin{eqnarray}
\label{SM Quantum H2}
&&\hat{H} =\sum_{\vec{\mathbf{k}},\alpha}^{}\hbar \omega ( \vec{\mathbf{k}} ) 
\hat{a}_{\vec{\mathbf{k}},\alpha}^{\dagger}
\hat{a}_{\vec{\mathbf{k}},\alpha}
\notag
\\  
&&+\frac{1}{2(\sqrt{2\pi})^3 }\epsilon_0 \int \mathrm{d}^3r \mathrm{d}^2q_p \mathrm{d}\nu_p \left \{ \frac{-1}{V} 
\sum_{\vec{\mathbf{k}}_1,\alpha_1;\vec{\mathbf{k}}_2,\alpha_2}
\tilde{\chi}_{ijl}^{(2)}\left[\vec{\mathbf{r}};~\omega(\vec{\mathbf{k}}_p),\omega(\vec{\mathbf{k}}_1),\omega(\vec{\mathbf{k}}_2)\right]
(\vec{\mathbf{\epsilon}}_{{\vec{\mathbf{k}}_p}})_{i}
(\vec{\mathbf{\epsilon}}_{{\vec{\mathbf{k}}_1,\alpha_1}})_{j}
(\vec{\mathbf{\epsilon}}_{{\vec{\mathbf{k}}_2,\alpha_2}})_{l}\right.
\notag
\\ 
&&\left .
\times \sqrt{\frac{\hbar^{2}\omega(\vec{\mathbf{k}}_1)\omega(\vec{\mathbf{k}}_2)}{4\epsilon_{0}^{2}} }
 ~e^{-i(\triangle \vec{\mathbf{q}} ) \cdot \vec{\mathbf{r}} } 
e^{-i\triangle k_z  \cdot z } e^{-i\triangle \omega  \cdot t } 
\tilde{E} _i(\vec{\mathbf{q}},\nu_p)
\hat{a}_{\vec{\mathbf{k}}_1,\alpha_1}^{\dagger}(t)
\hat{a}_{\vec{\mathbf{k}}_2,\alpha_2}^{\dagger}(t)
+ \mathrm{h.c.}
\right \}.
\end{eqnarray}
Here, $\triangle \vec{\mathbf{q}} \equiv  \vec{\mathbf{q}}_1 + \vec{\mathbf{q}}_2 - \vec{\mathbf{q}}_p$ denotes the transverse momentum mismatch, $\triangle k_z \equiv  k_{1z} + k_{2z} - k_{pz}$ denotes the longitudinal momentum mismatch, and $\triangle \omega \equiv \omega_1 + \omega_2 - \omega_p$ denotes the energy mismatch.

The nonlinear crystal is assumed to have dimensions $L_x$, $L_y$ and $L_z$, and to be centered at the origin $\vec{\mathbf{r}} = 0$. The spatial integration in Eq.\,\ref{SM Quantum H2} ranges from $-\frac{L_m}{2}$ to $+\frac{L_m}{2}$ (m=x,y,z) along each axis.
The crystal is assumed to be homogeneous in the transverse directions, such that the second-order nonlinear susceptibility $\tilde{\chi}_{ijl}^{(2)}$ is independent of the transverse coordinates. In our experimental implementation, the crystal is custom-poled along the $z$-axis with a poling profile $g(z)$ to engineer the desired spectral phase-matching function---specifically, to generate spectrally uncorrelated biphoton states. Consequently, after the integration over the entire space, the quantum Hamiltonian cold be written as:   
\begin{align}
\label{SM Quantum H3}
\hat{H} = & \sum_{\vec{\mathbf{k}},\alpha}
\hbar \omega ( \vec{\mathbf{k}} )
\hat{a}_{\vec{\mathbf{k}},\alpha}^{\dagger}
\hat{a}_{\vec{\mathbf{k}},\alpha}
\notag
\\ &  
+\frac{1}{2(\sqrt{2\pi})^3 }\epsilon_0
\int  \mathrm{d}^2 q_p \mathrm{d}\nu_p
\Bigg[
\frac{-L_x L_y}{V} 
\sum_{\vec{\mathbf{k}}_1,\alpha_1}
\sum_{\vec{\mathbf{k}}_2,\alpha_2}
\tilde{\chi}_{ijl}^{(2)}
(\omega(\vec{\mathbf{k}}_p),
\omega(\vec{\mathbf{k}}_1),
\omega(\vec{\mathbf{k}}_2))
(\vec{\mathbf{\epsilon}}_{\vec{\mathbf{k}}_p})_{i}
(\vec{\mathbf{\epsilon}}_{\vec{\mathbf{k}}_1,\alpha_1})_{j}
(\vec{\mathbf{\epsilon}}_{\vec{\mathbf{k}}_2,\alpha_2})_{l}
\notag
\\ &
\times
\sqrt{\frac{\hbar^{2}\omega(\vec{\mathbf{k}}_1)\omega(\vec{\mathbf{k}}_2)}{4\epsilon_{0}^{2}} }
~\mathrm{sinc}\!\left( \frac{\triangle q_x L_x}{2} \right)
\mathrm{sinc}\!\left( \frac{\triangle q_y L_y}{2} \right)
\left( \int \mathrm{d}z\, g(z)\, e^{-i\triangle k_z z} \right)
e^{-i\triangle \omega t}
\notag
\\ &
\times
\tilde{E}_i(\vec{\mathbf{q}},\nu_p)
\hat{a}_{\vec{\mathbf{k}}_1,\alpha_1}^{\dagger}(t)
\hat{a}_{\vec{\mathbf{k}}_2,\alpha_2}^{\dagger}(t)
+ \mathrm{h.c.}
\Bigg] .
\end{align}

Since the crystal dimensions are much larger than the optical wavelength, the discrete mode sum can be replaced by a continuum integral:
\begin{align}
\label{SM lim}
\lim_{V \to \infty} \frac{1}{V}\sum_{\vec{\mathbf{k}},\alpha} = 
\frac{1}{(2\pi)^3} \sum_{\alpha} \int \mathrm{d}^3k .
\end{align}

Further, with the polarization of signal and idler photons assumed to be fixed, the sums over $s_1$ and $s_2$ could be omitted. Defining the effective nonlinear coefficient $d_{\mathrm{eff} }\equiv \frac{1}{2} \tilde{\chi}_{ijl}^{(2)} $, the nonlinear part of the Hamiltonian simplifies to: 
\begin{align}
\label{SM Quantum H_NL}
H_{NL} \approx & C_{NL} d_{\mathrm{eff}} \iint \mathrm{d}^3k_1 \mathrm{d}^3k_2
\sqrt{\omega(\vec{\mathbf{k}}_1) \omega(\vec{\mathbf{k}}_2)}
\notag 
\\ &
\times \int \mathrm{d}^2 q_p \mathrm{d} \nu_p \left [ \prod_{m=1}^{2} \mathrm{sinc}\left ( \frac{\triangle k_m L_m}{2}  \right ) 
\times \int_{-\frac{L}{2}}^{\frac{L}{2}} \mathrm{d}z g(z) e^{-i\triangle k_z  \cdot z } \right ] 
\tilde{E} _i(\vec{\mathbf{q}},\nu_p)  e^{-i\triangle \omega  \cdot t }
\hat{a}^{\dagger}(\vec{\mathbf{k}}_1) \hat{a}^{\dagger}(\vec{\mathbf{k}}_2) + \mathrm{h.c.} .
\end{align}
Note that $C_{NL}$ is normalization constant.

Using first-order time-dependent perturbation theory, the biphoton state generated via SPDC is given by
\begin{align}
\label{SM Psi 1}
\left | \Psi (t)  \right \rangle =\left ( 1-\frac{i}{\hbar} \int_{0}^{t} \mathrm{d} t^{'}H_{NL}(t^{'}) \right ) \left | \Psi (0)  \right \rangle .
\end{align}

Since the vacuum state is annihilated by the negative-frequency terms, only the creation terms contribute. Assuming that the pump requires a time $T$ to propagate through the nonlinear crystal, the generated biphoton state can be written as:
\begin{align}
\label{SM Psi 2}
\left | \Psi_\mathrm{SPDC} \right \rangle & \approx  C_{0}\left | 0_{1},~0_{2} \right \rangle 
\notag 
\\ & 
+ C_{1}d_{\mathrm{eff}}  \iint \mathrm{d}^{3}k_1\mathrm{d}^3k_2\psi \left ( \vec{\mathbf{k}}_1,\vec{\mathbf{k}}_2 \right )  \sqrt{\omega_1( \vec{\mathbf{k}}_1 )\omega_2( \vec{\mathbf{k}}_2 )  }
\notag 
\\ &
\times e^{\frac{i\triangle \omega T}{2}}  \mathrm{sinc} \left ( \frac{\triangle \omega T}{2} \right )  \hat{a}^{\dagger} ( \vec{\mathbf{k}}_1 ) \hat{a}^{\dagger} ( \vec{\mathbf{k}}_2  )\left | 0_{1},~0_{2} \right \rangle , 
\end{align} 
and the joint spatial-spectral amplitude is gievn by:
\begin{align}
\label{SM JSSA 1}
\psi \left ( \vec{\mathbf{k}}_1,\vec{\mathbf{k}}_2 \right ) = \mathcal{N}  \int \mathrm{d}^2 q_p \mathrm{d} \nu_p 
\left [ \prod_{m=1}^{2} \mathrm{sinc}\left ( \frac{\triangle k_m L_m}{2}  \right ) 
\times \int_{-\frac{L}{2}}^{\frac{L}{2}} \mathrm{d}z g(z) e^{-i\triangle k_z  \cdot z }\right ]
\tilde{E} _i(\vec{\mathbf{q}}_p,\nu_p) .
\end{align}
Here, $\mathcal{N}$ is a normalization constant. We briefly explain the equivalence between a function expressed in terms of the wave vector $\vec{\mathbf{k}}$ and one expressed in terms of the transverse momentum $\vec{\mathbf{q}}$ and relative frequency $\nu$. For a given relative frequency $\nu$ and known central frequency $\Omega$, the magnitude of the wave vector $\left|\vec{\mathbf{k}}\right|$ is determined by the Sellmeier equation. Once the transverse momentum $\vec{\mathbf{q}}$ is specified, the longitudinal component follows as $k_z = \sqrt{\left | \vec{\mathbf{k}} \right | ^2 - \left | \vec{\mathbf{q}} \right | ^2 } $, which uniquely determines the wave vector $\vec{\mathbf{k}}$.  

To further simplify the expression, we note that the transverse dimensions of the crystal are much larger than the optical wavelength. Under this condition, the corresponding sinc functions can be approximated by Dirac delta functions, enforcing transverse momentum conservation, $\vec{\mathbf{q}}_1 + \vec{\mathbf{q}}_2 = \vec{\mathbf{q}}_p$. Similarly, when the nonlinear interaction time $T$ is much longer than the pump pulse duration, the temporal sinc function can be approximated by a delta function, enforcing energy conservation, $\omega_1 + \omega_2 = \omega_p$. For small transverse angles, the longitudinal phase mismatch can be approximated as $\triangle k_z \approx -\frac{\left | \vec{\mathbf{q}}_1 - \vec{\mathbf{q}}_2 \right |^2}{2k_p}$ which introduces an explicit dependence on the transverse momentum difference $\vec{\mathbf{q}}_1 - \vec{\mathbf{q}}_2$, and likewise on the frequency difference $\nu_1 - \nu_2$. Under these approximations, the biphoton joint spatial--spectral wavefunction factorizes into a product of a pump-envelope function and a phase-matching function:
\begin{equation}
\label{SM JSSA 2}
\psi( \vec{\mathbf{k}}_1,\vec{\mathbf{k}}_2) = \mathcal{N} \Phi \left ( \vec{\mathbf{q}}_1-\vec{\mathbf{q}}_2, \nu_1-\nu_2 \right )    \tilde{E}\left ( \vec{\mathbf{q}}_1+\vec{\mathbf{q}}_2,\nu_1+\nu_2 \right ),
\end{equation}
Here, $\Phi$ denotes the spatial--spectral phase-matching function and $\tilde{E}$ contains the spatial--spectral structure of the pump pulse. Therefore, the joint spatiotemporal wave function of the biphoton state is shaped not only by the phase-matching function of the nonlinear crystal, but also by the pump pulse spatial--spectral structure, which has often been overlooked in the previous work.

At this stage, we simplify the pump pulse as a direct product of a spatial $\mathrm{LG}_{00}$ mode and a spectral gaussian function with strong dispersion. We then discuss the resulting spatiotemporal correlations of the biphoton state and give a brief physical interpretation of their origin.
In the spatial domain, the longitudinal momentum of the pump pulse dominates, and its transverse momentum can be neglected, $\vec{\mathbf{q}}_p \approx 0$. Under transverse momentum conservation, $\vec{\mathbf{q}}_1 + \vec{\mathbf{q}}_2 = \vec{\mathbf{q}}_p \approx 0$, the signal and idler photons are therefore expected to exhibit anti-correlations in the transverse momentum basis.
In the spectral domain, the joint spectral phase is inherited from the dispersion of the pump pulse and can be expressed as a polynomial expansion in the relative pump frequency $\nu_p$, $\phi_{p} = c_0 + c_1\nu_p + c_2\nu_p^2 + c_3\nu_p^3+...$. The coefficient $c_0$ denotes a global phase, $c_1$ reflects the temporal arrival time of the pump pulse, while $c_2$, $c_3$, and higher-order coefficients describe the Group-Delay Dispersion\,(GDD), Third-Order Dispersion\,(TOD) and higher-order dispersion of the pump pulse, respectively. Due to the energy conservation, $\nu_p = \nu_1 + \nu_2$, thus the biphoton joint spectral phase could be approximated as $\phi_{\vec{\mathbf{q}}_1, \vec{\mathbf{q}}_2}(\nu_1,\nu_2) \approx c_0 + c_1(\nu_1 + \nu_2) + c_2(\nu_1 + \nu_2)^2 + c_3(\nu_1 + \nu_2)^3 + ...$. The extended cross terms, such as $2 c_2 \nu_1 \nu_2$, explicitly encode spectral phase correlations between the signal and idler photons. We emphasize that, although the KTP crystal is specifically engineered to generate spectrally uncorrelated biphoton state, the joint spectral phase remains sensitive to the dispersion of the pump pulse.
We further note that, following their generation, the individual free-space propagation of the signal and idler photons can independently modify the joint spectral phase through local operations, leading to additional separable spectral phases such as $c_{20} \nu_0^{2}$ and $c_{21} \nu_1^{2}$, which correspond to the dispersion accumulated by the signal and idler photons, respectively, during propagation.
%We note that due to the individual free-space propagation after generation, the joint spectral phase may further be manipulated by local operations, resulting in additive local spectral phase such as $c_{20} \nu_0^{2}$ and $c_{20} \nu_0^{2}$, corresponding to the local dispersion either signal or idler photons experienced during free-space propagation.  
%
Additionally, due to the intrinsic coupling between the transverse momentum $\vec{\mathbf{q}}$ and relative frequency $\nu$ in the phase-matching function, spatiotemporal correlations also arise within the individual wave packets of either the signal or idler photons. During free-space propagation following their generation, these spatiotemporal correlations can be further modified by optical aberrations of lenses, spatial angular dispersion introduced by prisms or gratings, and other system imperfections. As a result, our experimental observations provide primarily qualitative insight into the underlying spatiotemporal structure, and a more complete quantitative description will require further investigation.

%In addition, The poling function $g(z)$ is engineered to approximate a Gaussian phase-matching envelope.

\section{Spatially Resolved Joint Spectral Measurement}
While the joint intensity distribution of a biphoton state can be measured directly through simultaneous post-selection of both photons, characterization of the joint phase distribution requires a well-defined phase reference. To access the full phase correlations across the combined spatial–spectral degrees of freedom, we perform both spatially resolved joint spectral measurements and spectrally resolved joint spatial measurements. The former probes phase variations along the spatial dimension and is described in the Methods. In this section, we provide a detailed description of the spatially resolved joint spectral measurement.

\begin{comment}
In this section, we detail the method of the joint spatiotemporal measurement. While the joint intensity distribution of the biphoton state can be measured directly by performing post-selection simultaneously at both particles, the characterization of joint phase distribution requires a well-defined reference. To access the full phase correlations across the spatial-spectral degrees of freedom, we perform the spatially resolved joint spectral measurement and spectrally resolved joint spatial measurement. The former contains phase variation along the spectral dimension, and the latter contains that along the transverse spatial dimensions.
\end{comment}

To peform the spatially resolved joint spectral measurement, we employ a spatially resolved spectral shearing interferometer (SSI). In this scheme, the idler photons are directly post-selected in both the spatial and spectral domains. Meanwhile, the signal photons are routed through a Mach–Zehnder interferometer (MZI), where one arm applies a spectral shear of $\Omega_s$, and the other introduces a temporal delay $\tau_s$. At the output of the MZI, the signal photons are also post-selected in the spatial domain, and their spectral interference patterns are measured. The resulting interference pattern can be expressed as:
\begin{equation}
    \label{JSA M1}
    \begin{aligned}
&S_{\vec{\mathbf{q}}_s,\vec{\mathbf{q}}_i,\Omega_s,\tau_s} \left ( \omega_{s},\omega_{i} \right ) 
\\=& \left |\psi_{\vec{\mathbf{q}}_s,\vec{\mathbf{q}}_i} \left ( \omega_{s}, \omega_{i}\right )e^{i\omega_{s}\tau_s} + \psi_{\vec{\mathbf{q}}_s,\vec{\mathbf{q}}_i} \left ( \omega_{s} +  \Omega_s , \omega_{i}\right )  \right |^{2} 
\\=& S_{\vec{\mathbf{q}}_s,\vec{\mathbf{q}}_i} \left ( \omega_{s}, \omega_{i}\right ) + S_{\vec{\mathbf{q}}_s,\vec{\mathbf{q}}_i} \left ( \omega_{s} + \Omega_s , \omega_{i}\right )  
\\
& + 2\mathrm{Re}\left [\psi_{\vec{\mathbf{q}}_s,\vec{\mathbf{q}}_i} \left ( \omega_{s}, \omega_{i}\right )\psi_{\vec{\mathbf{q}}_s,\vec{\mathbf{q}}_i}^{\ast } \left ( \omega_{s} + \Omega , \omega_{i}\right ) e^{i\omega_{s}\tau_s} \right ] .
    \end{aligned}
\end{equation}

To extract the joint spectral phase, we perform a Fourier transform along the signal frequency axis of the measured interference pattern $S_{\vec{\mathbf{q}}_s,\vec{\mathbf{q}}_i,\Omega_s,\tau_s} \left ( \omega_{s},\omega_{i} \right )$, yielding:
\begin{equation}
    \label{JSA M2}
    \begin{aligned}
\tilde{S}_{\vec{\mathbf{q}}_s,\vec{\mathbf{q}}_i,\Omega_s,\tau_s} \left ( T_{s},\omega_{i} \right ) 
= & \mathcal{F} \left [ S_{\vec{\mathbf{q}}_s,\vec{\mathbf{q}}_i,\Omega_s,\tau_s} \left ( \omega_{s},\omega_{i} \right )  \right ]
\\ = & \frac{1}{2} \tilde{S}_{\vec{\mathbf{q}}_s,\vec{\mathbf{q}}_i} \left ( T_{s},\omega_{i} \right ) \left ( 1 + e^{iT_s\Omega_s} \right )
\\ + & \mathcal{F}\left \{ \left | \psi_{\vec{\mathbf{q}}_s,\vec{\mathbf{q}}_i} \left ( \omega_{s}, \omega_{i}\right ) \psi_{\vec{\mathbf{q}}_s,\vec{\mathbf{q}}_i}^{*} \left ( \omega_{s}+\Omega_s, \omega_{i}\right ) \right | e^{i\triangle \phi\left ( \omega_s,\Omega_s,\omega_i \right ) } \right \}
 \circledast \delta\left ( T-\tau_s  \right )  
\\ + & \mathcal{F}\left \{ \left | \psi_{\vec{\mathbf{q}}_s,\vec{\mathbf{q}}_i}^{*} \left ( \omega_{s}, \omega_{i}\right ) \psi_{\vec{\mathbf{q}}_s,\vec{\mathbf{q}}_i} \left ( \omega_{s}+\Omega_s, \omega_{i}\right ) \right | e^{-i\triangle \phi\left ( \omega_s,\Omega_s,\omega_i \right ) } \right \}
 \circledast \delta\left ( T+\tau_s  \right ) .
    \end{aligned}
\end{equation}
Here, $\circledast$ denotes the convolution operator, and phase difference is defined as $\triangle\phi\left ( \omega_s,\Omega_s,\omega_i \right )  =\phi_{\vec{\mathbf{q}}_s,\vec{\mathbf{q}}_i} \left ( \omega_{s}, \omega_{i}\right ) - \phi_{\vec{\mathbf{q}}_s,\vec{\mathbf{q}}_i} \left ( \omega_{s}+\Omega_s, \omega_{i}\right )$, which encodes the joint spectral phase gradient information along the signal axis.

If the time interval $\tau_s$ is sufficiently large, the three terms above become temporally well-separated---centered around $\pm \tau_s$ and 0, respectively. By applying a window function centered at $T_s = +\tau_s$, and performing an inverse Fourier transform, one isolates the interference term to obtain:
\begin{equation}
    \label{JSA M3}
    \begin{aligned}
\left | \psi_{\vec{\mathbf{q}}_s,\vec{\mathbf{q}}_i} \left ( \omega_{s}, \omega_{i}\right ) \psi_{\vec{\mathbf{q}}_s,\vec{\mathbf{q}}_i}^{*} \left ( \omega_{s}+\Omega_s, \omega_{i}\right ) \right | e^{i\left [ \omega_s\tau_s + \triangle\phi\left ( \omega_s,\Omega_s,\omega \right ) \right ] } .
    \end{aligned}
\end{equation}
Once the parameters $\Omega_s$ and $\tau_s$ have been calibrated, the argument of the exponential yields the quantity $\theta_{\vec{\mathbf{q}}_s,\vec{\mathbf{q}}_i}\left ( \omega_s,\Omega_s,\tau_s,\omega_i \right ) = \omega_s \tau_s + \triangle\phi\left ( \omega_s,\Omega_s,\omega_i \right )$, and the joint spectral phase along the signal axis is then obtained via integration:
\begin{equation}
\label{JSA M4}
\phi_{\vec{\mathbf{q}}_s,\vec{\mathbf{q}}_i} \left ( \omega_{s}, \omega_{i}\right ) = \frac{1}{\Omega_s} \int \left [ \tau_s\delta \omega_s - \theta_{\vec{\mathbf{q}}_s,\vec{\mathbf{q}}_i}\left ( \omega_s,\Omega_s,\tau_s,\omega_i \right )  \right ]\mathrm{d}\omega_s .
\end{equation}

However, the relative spectral phase between different idler frequencies is undetermined. We exchange the paths of the signal and idler photons and repeat the experiments to measure the joint spectral phase gradient along the idler axis $\triangle\phi\left ( \omega_i,\Omega_i,\omega_s \right )  =\phi_{\vec{\mathbf{q}}_s,\vec{\mathbf{q}}_i} \left ( \omega_{s}, \omega_{i}\right ) - \phi_{\vec{\mathbf{q}}_s,\vec{\mathbf{q}}_i} \left ( \omega_{s}, \omega_{i}+\Omega_i\right )$. Once the phase gradients along both the signal and idler frequency axes are obtained, the spatially resolved joint spectral phase $\phi_{\vec{\mathbf{q}}_s,\vec{\mathbf{q}}_i} \left ( \omega_{s}, \omega_{i}\right )$ could be completely reconstructed by zonal approach\,\cite{Southwell1980}.

\bibliography{references}